\documentclass[aps,prx,reprint,superscriptaddress]{revtex4-2}

\usepackage{graphicx}[pdflatex]
\usepackage{dsfont}
\usepackage{amsmath}
\usepackage{xcolor} 
\usepackage{bbold}
\usepackage[normalem]{ulem}

\begin{document}

\title{A M{\o}lmer-S{\o}rensen Gate with Rydberg-Dressed Atoms}


\author{Michael J. Martin}
\affiliation{Sandia National Laboratories, Albuquerque, New Mexico 87123, USA}
\affiliation{Los Alamos National Laboratory, Los Alamos, New Mexico 87545, USA}
\affiliation{Center for Quantum Information and Control, University of New Mexico, Albuquerque, New Mexico 87131, USA}
\affiliation{Department of Physics and Astronomy, University of New Mexico, Albuquerque, New Mexico 87106, USA}

\author{Yuan-Yu Jau}
\affiliation{Sandia National Laboratories, Albuquerque, New Mexico 87123, USA}
\affiliation{Center for Quantum Information and Control, University of New Mexico, Albuquerque, New Mexico 87131, USA}
\affiliation{Department of Physics and Astronomy, University of New Mexico, Albuquerque, New Mexico 87106, USA}

\author{Jongmin Lee}
\affiliation{Sandia National Laboratories, Albuquerque, New Mexico 87123, USA}
\affiliation{Center for Quantum Information and Control, University of New Mexico, Albuquerque, New Mexico 87131, USA}
\affiliation{Department of Physics and Astronomy, University of New Mexico, Albuquerque, New Mexico 87106, USA}

\author{Anupam Mitra}
\affiliation{Center for Quantum Information and Control, University of New Mexico, Albuquerque, New Mexico 87131, USA}
\affiliation{Department of Physics and Astronomy, University of New Mexico, Albuquerque, New Mexico 87106, USA}

\author{Ivan H. Deutsch}
\affiliation{Center for Quantum Information and Control, University of New Mexico, Albuquerque, New Mexico 87131, USA}
\affiliation{Department of Physics and Astronomy, University of New Mexico, Albuquerque, New Mexico 87106, USA}

\author{Grant W. Biedermann}
\affiliation{Sandia National Laboratories, Albuquerque, New Mexico 87123, USA}
\affiliation{Department of Physics and Astronomy, University of Oklahoma, Norman, Oklahoma 73019, USA}
\affiliation{Center for Quantum Information and Control, University of New Mexico, Albuquerque, New Mexico 87131, USA}
\affiliation{Department of Physics and Astronomy, University of New Mexico, Albuquerque, New Mexico 87106, USA}


\date{\today}

\begin{abstract}
\noindent Neutral atoms are building blocks of ground-up quantum many-body systems. Well-controlled and high-fidelity entangling gates are an essential component for realizing complex neutral atom architectures for quantum computing, quantum simulation, and measurement with precision better than the standard quantum limit. In this Letter we report the realization of a M{\o}lmer-S{\o}rensen unitary between two neutral atoms, based on adiabatic single-photon dressing to Rydberg levels. We show that this technique is highly robust to noise sources and experimental imperfections that have limited the fidelity of other approaches to neutral atom gates. 
\end{abstract}


\maketitle

\section{Introduction \label{Introduction}}
Neutral atom systems that employ excitation of single atoms to Rydberg states are emerging as a promising resource for quantum computing \cite{Jaksch2000, Saffman2010, Wilk2010, Zhang2010, weiss2017quantum, Levine2019, madjarov2020high, henriet2020quantum}, quantum simulation \cite{Glaetzle2015, labuhn2016tunable, Bernien2017, celi2020emerging, ebadi2021quantum}, and entanglement-enhanced sensing \cite{Gil2014, brif2020characterization}. In particular, the strong van der Waals interaction between Rydberg levels is a resource for performing entangling operations between two or more atoms. 
A decade after the first demonstration of an entanglement between neutral atoms via the Rydberg blockade \cite{Wilk2010, Zhang2010}, recent advances have demonstrated gates with entanglement fidelities as high as 97\% \cite{levine2018high, Levine2019}, Rydberg blockade-induced Bell-state preparation with fidelity $> 99.1$\% \cite{madjarov2020high}, and gate operations within arrays of 121 sites with controlled-Z gate fidelity of  89\% \cite{graham2019rydberg}.

These recent advances in entangling gates are limited by three problems that pose technical challenges for improving gate fidelity. First, approaches that employ two-photon schemes as in Refs. \cite{Levine2019, graham2019rydberg, madjarov2020high} 
are susceptible to off-resonant scattering from the intermediate state, as well as differential light shifts on the qubit states that must be accounted for. Second, as a general problem, thermal motion of the atoms causes dephasing during an entangling gate operation \cite{Wilk2010, Zhang2010, Keating2015, Leseleuc2018}. Third, schemes that require single atom-resolved excitation to Rydberg states  are susceptible to inhomogeneities in coupling strength to the Rydberg state \cite{graham2019rydberg}. This inhomogeneity can be worsened by strong intensity gradients of tightly-focused excitation lasers coupled to atomic position fluctuations and beam pointing instability. Some experimentally-demonstrated gate schemes have addressed at most one of these restrictions, \textit{e.g.}, as demonstrated by Levine \textit{et al.} Ref.~ \cite{Levine2019}, where a geometric phase-based approach allowed simultaneous Rydberg excitation of two or more atoms as part of the gate, which permitted larger beam sizes, reducing misalignment sensitivity. Direct, single-photon coupling to the Rydberg state solves the problem of scattering and light shifts originating from the intermediate level \cite{Hankin2014}. Further, Rydberg-dressed states \cite{Bouchoule2002spin,  Johnson2010, honer2010collective, henkel2010three, Keating2015, Jau2016} provide a powerful new setting for adiabatic controlled-phase gates \cite{Keating2015}. Previous experimental work with single photon-coupled, Rydberg-dressed states demonstrated a spin-flip blockade, allowing the preparation of entangled Bell states \cite{Jau2016}. However, while this work demonstrated the power of the dressed-atom Rydberg blockade effect, it did not demonstrate a protocol that was capable of acting on all possible logical input states. The situation is analogous to works that demonstrate Bell state production via Rydberg blockade \cite{Wilk2010, Zhang2010, madjarov2020high}, which are not sufficient for a universal gate set.

In this Letter, we experimentally demonstrate the protocol analyzed in Ref. \cite{Mitra2020} that addresses all of these challenges: a neutral atom M\o{}lmer-S\o{}rensen (MS) gate. We implement a controlled phase gate between Rydberg-dressed atoms that is embedded within a spin-echo topology \cite{Mitra2020}. This realizes a unitary operation in the two-qubit basis that maps onto the M\o{}lmer-S\o{}rensen gate \cite{molmer1999multiparticle, sorensen1999quantum}, given by $\hat{U}_{\mathrm{MS}} = \exp \left[-i \phi_{J} \hat{S}_{y}^{2} \right]$. Our approach combines the benefits of Rydberg dressing with a greatly reduced sensitivity to  atomic motion. 
\begin{figure*}[t]
 \includegraphics[width=2 \columnwidth]{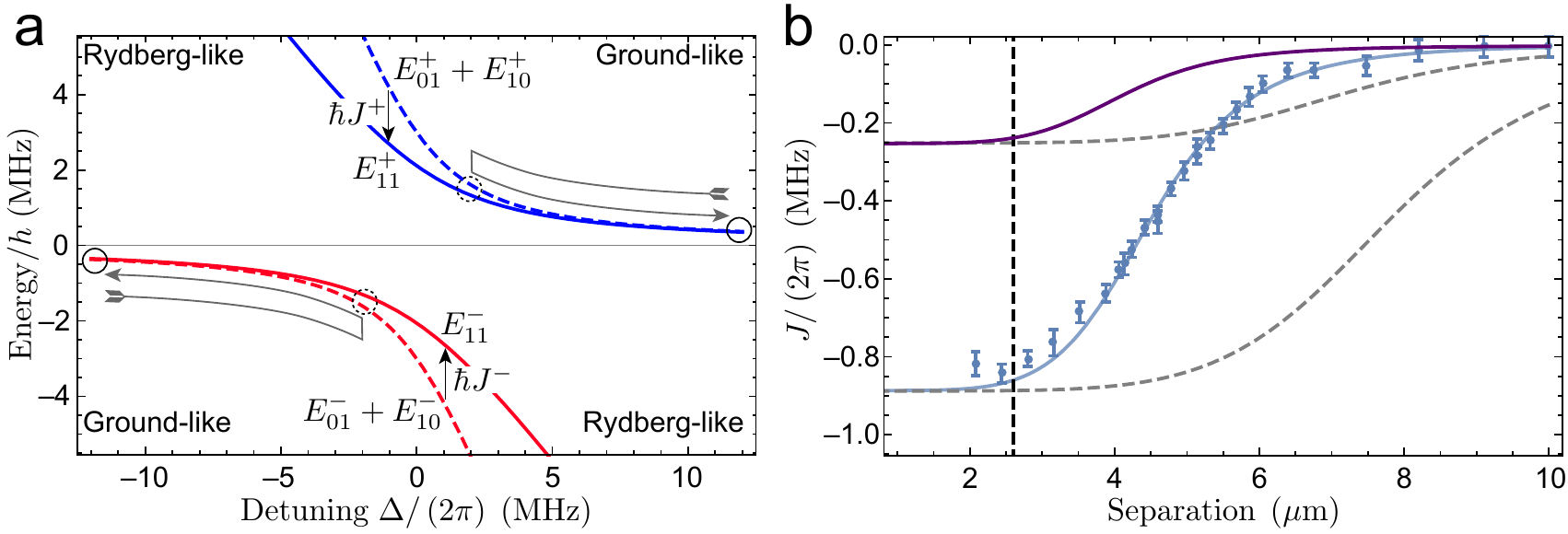}%
 \caption{\label{JCurve} (a) Energy eigenvalues of the light shift Hamiltonian with (solid) and without (dashed) Rydberg interaction. The difference between the two cases yields the dressed interaction energy, $J^{\pm}$, where the $+$ ($-$) indicates the higher (lower) energy branch of the eigenspectrum. Here, $\Omega=2 \pi \times 2.95$~ MHz, and $r=2.6$~$\mu$m, which correspond to the values used in the experiment. The inferred $c_{6}$ coefficient used in the calculation is $c_{6}=25\;\mathrm{GHz}\times\mu \mathrm{m}^{6}$. (b) Measured $J^{+}$ for $\Omega=2 \pi \times 2.95(5)$~MHz and $\Delta= -0.2(2)$~MHz (data points and fit).  These data were obtained via a dressed Ramsey scheme where atom~1 was kept in the $|1\rangle$ state, and atom~2 was independently manipulated. The fit to the curve yields $c_{6}=25(1)\;\mathrm{GHz}\times\mu \mathrm{m}^{6}$. The purple solid curve represents $J^{+}$ used in the entangling gate, with values for Rabi frequency and detuning used in the calculation fixed at $\Omega = 2 \pi \times 2.95$~MHz and $\Delta = 2 \pi \times 2.0$ MHz. The dashed curves represent the inter-atomic potential for the quantization axis perpendicular to the inter-atomic axis, whereas the solid curves the two axes are aligned, which was the condition we used in all experiments reported here. We note that in this configuration, the blockade radius $\left(c_{6}/\Omega\right)^{1/6}$, is larger by a factor of 1.7.}
 \end{figure*}
In this work, we consider pairs of three-level atoms. Two of the levels comprise qubit states in the ground hyperfine manifold, $|0\rangle \equiv |F=3, m_{f} =0\rangle$ and  $|1\rangle \equiv |F=4, m_{f} =0\rangle$. The third level is the Ryberg state $|R\rangle \equiv |64P_{3/2}, m_{j} =3/2\rangle$. The qubit manifold is coupled to the $|R\rangle$ state via a single-photon coupling with Rabi frequency $\Omega_{j}$ and detuning $\Delta_{j} = \omega_{l}-\omega^{j}_{R}$ for atom $j \in \{1,2\}$. The coupling is near-resonant with the $|1\rangle \rightarrow |R\rangle$ transition, which realizes the Hamiltonian $\hat{H} =   \sum_{j=1,2} \hat{H}_{0}^{j}  +\hat{H}_{VdW}$  where 
\begin{equation}
\hat{H}_{0}^{j}= \hbar \left( \frac{\Omega_{j}}{2} |1_{j}\rangle\langle R_{j}|+\frac{\Omega_{j}^*}{2} |R_{j}\rangle\langle 1_{j}|  -\Delta_{j} |R_{j}\rangle \langle R_{j}|\right)
\end{equation}
and  
\begin{equation}
\hat{H}_{\mathrm{VdW}} = - \left(c_{6}/r^{6}\right) |R_{1}R_{2}\rangle \langle R_{1} R_{2}|\mathrm{,}
\end{equation}
where $r$ is the separation between atoms.

\begin{figure}[b]
 \includegraphics[width=1 \columnwidth]{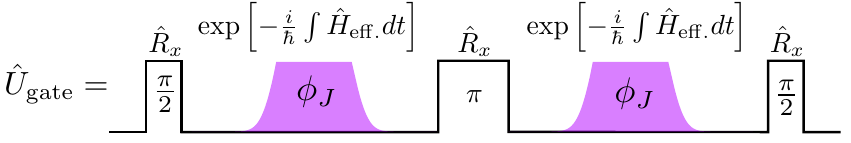}%
 \caption{\label{EchoSequence} Echo sequence showing the dressing operations that comprise $\hat{U}_{\mathrm{gate}}$. 
  }
 \end{figure}

Diagonalizing $\hat{H}$ yields the energy spectrum of the atom-light plus Van der Waals Hamiltonian. The effect of  $\hat{H}_{\mathrm{VdW}}$ modifies the energy of the $|\widetilde{11}\rangle$ state, where the tilde indicates dressing, such that $E_{11}^{\pm}= E_{01}^{\pm} + E_{10}^{\pm}+\hbar J_{\pm}\left(\Delta, \Omega\right)$, where $E_{10}^{\pm}$ ($E_{01}^{\pm}$) are the eigenvalues of $ \hat{H}_{0}^{1}$ ($ \hat{H}_{0}^{2}$), given by $\hbar \left(-\frac{1}{2} \Delta_{j} \pm \sqrt{|\Omega_{j}|^{2} +\Delta_{j}^{2}}\right)$, and the $+$ ($-$) sign refer to eigenstates that adiabatically map to ground states under large positive (negative) detuning, with corresponding positive (negative) energy eigenvalues of the dressed states. In the limit of small inhomogeneity, where $\overline{\Omega} \equiv \left(\Omega_{1} +\Omega_{2}\right)/2$ is much greater in magnitude than $|\Omega_{1}-\Omega_{2}|/2$ and $\overline{\Delta} \equiv \left(\Delta_{1} +\Delta_{2}\right)/2$ is much greater in magnitude than $|\Delta_{1}-\Delta_{2}|/2$, the light shift Hamiltonian in the dressed basis becomes
\begin{equation}
\hat{H}_{\mathrm{eff.}}^{\pm} =  E_{10}^{\pm} |\widetilde{1_{1}}\rangle \langle \widetilde{1_{1}}| +  E_{01}^{\pm}|\widetilde{1_{2}}\rangle \langle \widetilde{1_{2}}| + \hbar J_{\pm} |\widetilde{11}\rangle\langle\widetilde{11}|, \label{DressedHamiltonia}
\end{equation}
where in the limit that the magnitude of the van~der~Waals interaction, $|c_{6}/r^6|$, is much greater than either $|\overline{\Omega}|$ or $|\overline{\Delta}|$, 
\begin{equation}
J_{\pm} \rightarrow \frac{1}{2} \overline{\Delta} \pm\frac{1}{2} \left(\sqrt{2 |\overline{\Omega}|^{2}+\overline{\Delta}^{2}} - 2 \sqrt{|\overline{\Omega}|^{2} +\overline{\Delta}^{2}}\right).
\end{equation}
Figure~\ref{JCurve} shows the eiegenvalues of $\hat{H}$ and $\hat{H}_{0}$ for experimentally-relevant parameters, and Fig.~\ref{JCurve}b shows corresponding measurements of $J_{+}$ taken near $\overline{\Delta}=0$. Fitting the data in Fig.~\ref{JCurve}b to the numerical expression for $J_{+}$ yields an inferred $c_{6}$ coefficient of of $25(1)\;\mathrm{GHz} \times \mu \mathrm{m}^{6}$, which is in approximate agreement with $c_{6} = 27.5 \; \mathrm{GHz} \times \mu \mathrm{m}^{6}$, obtained via a perturbative calculation \cite{vsibalic2017arc}, with the inter-atomic axis aligned to the quantization axis (if perpendicular, $c_6$ will be approximately 25 times larger). We note that $ \hat{H}_{\mathrm{eff.}}^{+}$ and $\hat{H}_{\mathrm{eff.}}^{-}$, corresponding to different branches of the of the atom-light plus van~der~Waals Hamilitonian, may be each implemented via adiabatic preparation via dynamical control of $\overline{\Delta}$ and $\overline{\Omega}$ \cite{Lee2017}, the technique we employ here. We also note that the entangling energy is inherently robust with respect to inhomogeneity in Rabi frequency and detuning. In what follows, we consider the $J_{+}$ branch, as this is the branch we implement in the experiment, and thus omit the label. We also consider only the case where $\Omega_{1}=\Omega_{2}=\Omega$ and $\Delta_{1} = \Delta_{2}=\Delta$, thus $E_{10}=E_{01}\equiv E_{\mathrm{LS}}^{(1)}$ (as in Ref.~\cite{Mitra2020}). Further, embedding the operation of $\hat{H}_{\mathrm{eff.}}$ into a spin echo sequence as shown in Fig.~\ref{EchoSequence}, removes single atom terms and yields the two-qubit MS operator given by
\begin{equation} 
\hat{U}_\mathrm{gate} = \exp \left[- i \phi_{J}  \left(\hat{S}_{y}\right)^{2} \right],
\label{UGate}
\end{equation}
where $\hat{S}_y = \hat{s}_{y}^{1} + \hat{s}_{y}^{2}$, with $\hat{s}_{y}^{j} = i \left(|0_{j}\rangle \langle 1_{j}|-|1_{j}\rangle \langle 0_{j}|\right)/2$ 
and
\begin{equation}
\phi_{J} = \int J\left(t\right) dt.
\label{PhiJ}
\end{equation}
We note that Eqns.~\ref{UGate} and \ref{PhiJ} are only valid in the limit of adiabatic following of the light shift potentials, which sets a limit on the rate of change of the parameters $\Omega \left(t\right)$ and $\Delta \left(t\right)$. When $|\phi_{J}| = \pi/2$, $\hat{U}_{\mathrm{gate}}$ is a perfect entangler~\cite{zhang2003geometric}.

Our experiment  \cite{Hankin2014, Jau2016}  employs two individually-trapped ultracold cesium atoms confined in optical tweezers at the Cs D2 magic wavelength of 936 nm \cite{ye2008quantum}. The optical tweezers are formed by re-imaging an acousto-optic modulator that is driven with two rf tones onto a high numerical-aperture asphere, thus projecting two traps with tunable spacing and $1/e^{2}$ intensity radius consistent with $w_{0}=1.1$~$\mu$m in the trap focal plane. The two atoms are loaded from a background magneto-optical trap into the tweezers in the strong collisionally-blockaded regime \cite{schlosser2001sub}, with trap-resolved fluorescence detected by two separate avalanche photodiodes (APDs). The APD signal is used to detect the simultaneous presence of two atoms in the traps, as well as for state measurement.

After loading two atoms into two separate tweezers, a stage of polarization-gradient cooling is followed by optical pumping on the D1 $|F=4, m_{f} =0\rangle \rightarrow |F'=4, m'_{f} = 0\rangle$ transition to spin-polarize the atoms into the $|1\rangle \equiv |F=4, m_{f} =0\rangle$ qubit state with $96.3(8)\%$ state preparation efficiency. We observe that the imperfections arise due to population of other spin states in the $F=4$ manifold, consistent with polarization impurities or misalignment with respect to the quantization axis in our optical pumping stage. The $|1\rangle$ state is coupled to the $|0\rangle \equiv |F=3, m_{f} =0\rangle$ state by a 
Raman coupling scheme that is detuned from the D2 cycling transition by -80 GHz. Here, a first beam, with sidebands at 9.2~GHz produced with an electro-optic modulator (EOM), is combined with a second beam derived from the same Raman laser that has an additional frequency offset applied via an acousto-optic modulator. We realize a Rabi frequency ($\pi$-pulse time) of $\Omega_{\mu \mathrm{w}} = 2 \pi \times 410$~kHz (1.2~$\mu$s) for the $|0\rangle \rightarrow |1\rangle$ qubit transition. During a given experimental sequence, the Raman coupling light is left on, while the rf drive of the EOM is rapidly switched to produce pulses with precise and reproducible timing characteristics. 

State detection is achieved via a separate retro-reflected beam that is near resonant with the D2 $\vert 6S_{1/2}, F=4 \rangle \rightarrow \vert 6P_{3/2}, F'=5\rangle$ cycling transition. A knife-edge mirror separates the signal from atom~1 and atom~2 at a secondary image plane. When in the $F=4$ manifold, an average of 3.8 (2.9) photons are detected for Atom~1 (Atom~2).  After state sensitive detection, which yields either a ``bright'' $F=4$ state ($\geq1$ photon measured) or a ``dark'' $F=3$ state (no photons recorded). A final detection step pumps atoms from $F=3$ and confirms the presence of a ``bright'' state. A null measurement on this final check indicates that the atom has been lost, and the data is discarded. We note that this detection scheme can only discriminate between the entire $F=4$ and $F=3$ manifolds, and does not distinguish the $m_{F}=0$ qubit levels from population in other Zeeman states. This detection system allows rapid experimental duty cycle while both atoms remain, with several re-uses for each atom pair common. We further assess the detection fidelity of this approach by pumping the atom to a bright or dark state with high probability. For the results described below, we measured at the conclusion of the experiment that $P_{B1}=93.9(1)\%$  and $P_{B2}=90.8(1)\%$, where $P_{B1}$ ($P_{B2}$) is the probability that if atom~1 (atom~2) is prepared in the bright state, the state detection correctly identifies its state; and $P_{D1}=1.62(3)\%$ and $P_{D2}=4.56(5)\%$, the probability either atom 1 or atom 2, respectively ,is misidentified as a bright state when prepared in a dark state. Here, the errors quoted are statistical bounds based on the measurement alone. Because the experiment runs over 12 hours, we include a 2\% systematic uncertainty in the bright state detection efficiency, which can arise from fluctuations in experimental parameters that influence mean photons detected for an atom a bright state. We bound this systematic uncertainty by known experimental characteristics and time-dependent analysis of the data.
\begin{figure*}
\includegraphics[width=2 \columnwidth]{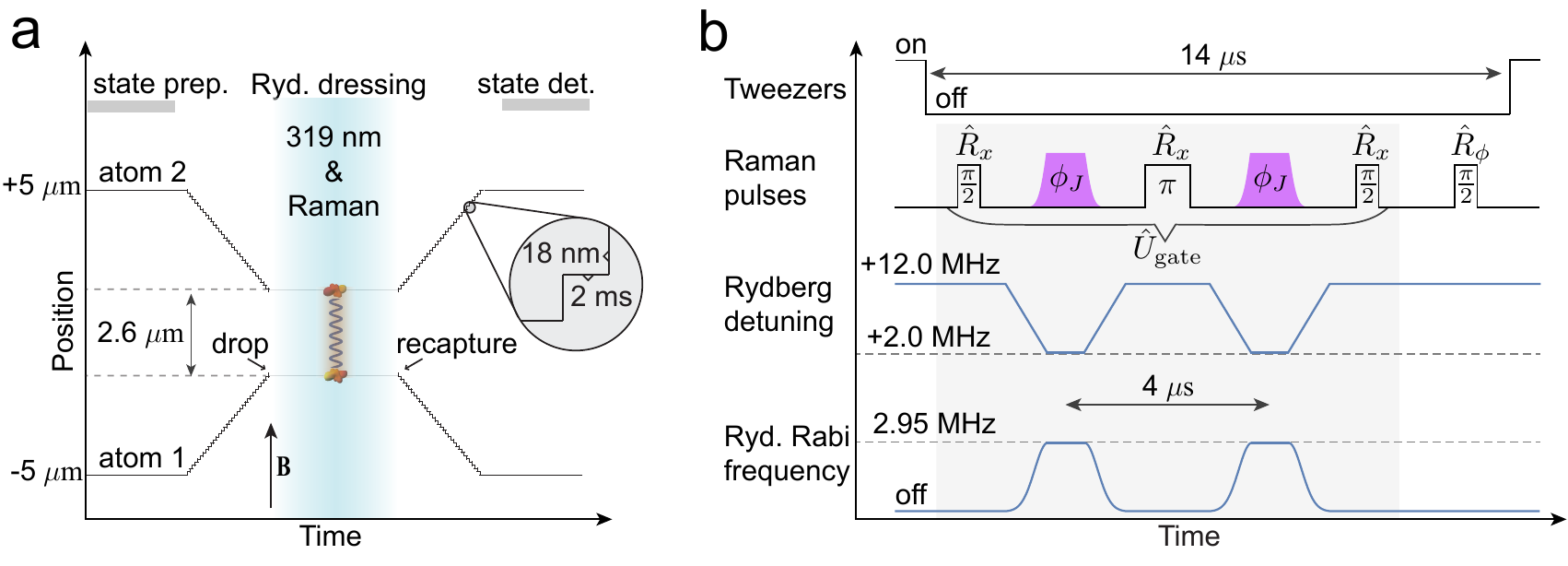}%
\caption{\label{Timing} (a) Experimental sequence to prepare, entangle and detect the atoms. Each atom is trapped in a separate optical tweezer. Both tweezers are translated so that the atoms are trapped within 2.6~$\mu$m of each other. The tweezers are shut off for Rydberg and Raman manipulations before the atoms are re-captured, brought apart, and their internal state is detected. (b) Timing diagram for the relevant operations in the experiment. The optical tweezers, Raman pulses, and Rydberg laser amplitude and detuning are dynamically controlled to impart the desired gate operation, while maintaining adiabaticity in the Rydberg dressing. A final $\pi/2$-pulse with rotation axis $\{\cos\left(\phi\right), \sin\left(\phi\right), 0\}$ enables characterization of the prepared entangled state.  
}
\end{figure*}

The $|1\rangle$ qubit state is coupled to the $|64P_{3/2}, m_{j} =3/2\rangle$ Rydberg level via a single-photon excitation scheme at 319~nm \cite{Hankin2014}. The 319~nm laser is focused onto the pair of atoms with waist $w_0=12$~$\mu$m $1/e^{2}$ and $\sigma^{+}$ polarization with respect to our bias field of 4.6~G. Maximum achievable powers are at the level of 50~mW. The bias field splits the $m_{j}$ sublevels of the  $64P_{3/2}$ Rydberg states by 8.6 MHz, clearly resolving the  $|64P_{3/2}, m_{j} =3/2\rangle$ state for Rydberg dressing. Further, by operating at positive detuning, we eliminate the possible influence of the $|64P_{3/2}, m_{j} =1/2\rangle$ state. We carefully control the electric field with a Faraday shield and 8 electrodes \cite{Lee2017} to within $\lesssim 10$~mV/cm of zero as this maximizes the efficiency of our Rydberg excitation and provides quadratic insensitivity to electric field fluctuations. We note that the light shift of the $|1\rangle \rightarrow |R \rangle$ resonance from the Raman light is included in our reported Rydberg detuning. For the results presented here, we operate the experiment such that $\Omega=2 \pi \times 2.95(5)$~MHz. With $\Delta=0$, we realize entangling energy scales of approximately 860~kHz, as shown in Fig.~\ref{JCurve}, with the expected dependence on inter-atomic separation and $c_{6}$~coefficient of the correct magnitude for our chosen quantization axis, which is aligned with the vector that connects the two atoms' centers of mass. While our entangling energy can be as large as $\sim 1$~MHz, we find that operating with $\Delta=+2\pi\times 2$~MHz has a less sensitive dependence on total gate time, which makes identifying the dressing time required produce to $\phi_{J} = - \pi/2$ more robust. A careful study of dressing parameters as they relate to entanglement fidelity is a topic for future work. In the following, we operate with $\Delta = 2 \pi \times 2.0(2)$~MHz, and $\Omega= 2 \pi \times 2.95(5)$~MHz. 

We entangle the atoms following the procedure operations detailed in Fig.~\ref{Timing}, which involves spatial manipulation (Fig.~\ref{Timing}a) and control of internal degrees of freedom and entanglement (Fig.~\ref{Timing}b). Specifically, after the initial cooling and state preparation, we translate the tweezers from their initial 10~$\mu$m separation to the final $2.6$~$\mu$m separation. Next, adiabatic ramping of the radial tweezer trap frequency (depth) from 55~kHz (600 $\mu\mathrm{K}$) to 34~kHz (200 $\mu\mathrm{K}$) cools the center of mass motion to an average temperature of approximately 10~$\mu$K. Next, the trap is abruptly switched off, and we apply the first $\pi/2$ pulse, followed by the dressed gate echo sequence of Fig.~\ref{Timing}b, indicated in the shaded box. We target a dressing phase $\phi_{J}=-\pi/2$ to realize a perfect entangler. For the data presented here, we measured that the dressing parameters of our sequence produced $\phi_{J} = -1.53(1)$, which represents a 2.5\% error relative to our target $\phi_{J}$. The impact of an error in $\phi_{J}$ at this level will impact the target state fidelity at the level of 0.1\%, and is thus negligible at our level of precision.  We follow the entangling operation by a final $\pi/2$ along the same axis as the preceding Raman pulses. In the case of a perfect sequence and entangling operation, we expect that the ideal output state will be a Bell state. 

\begin{figure}[t]
\includegraphics[width=1 \columnwidth]{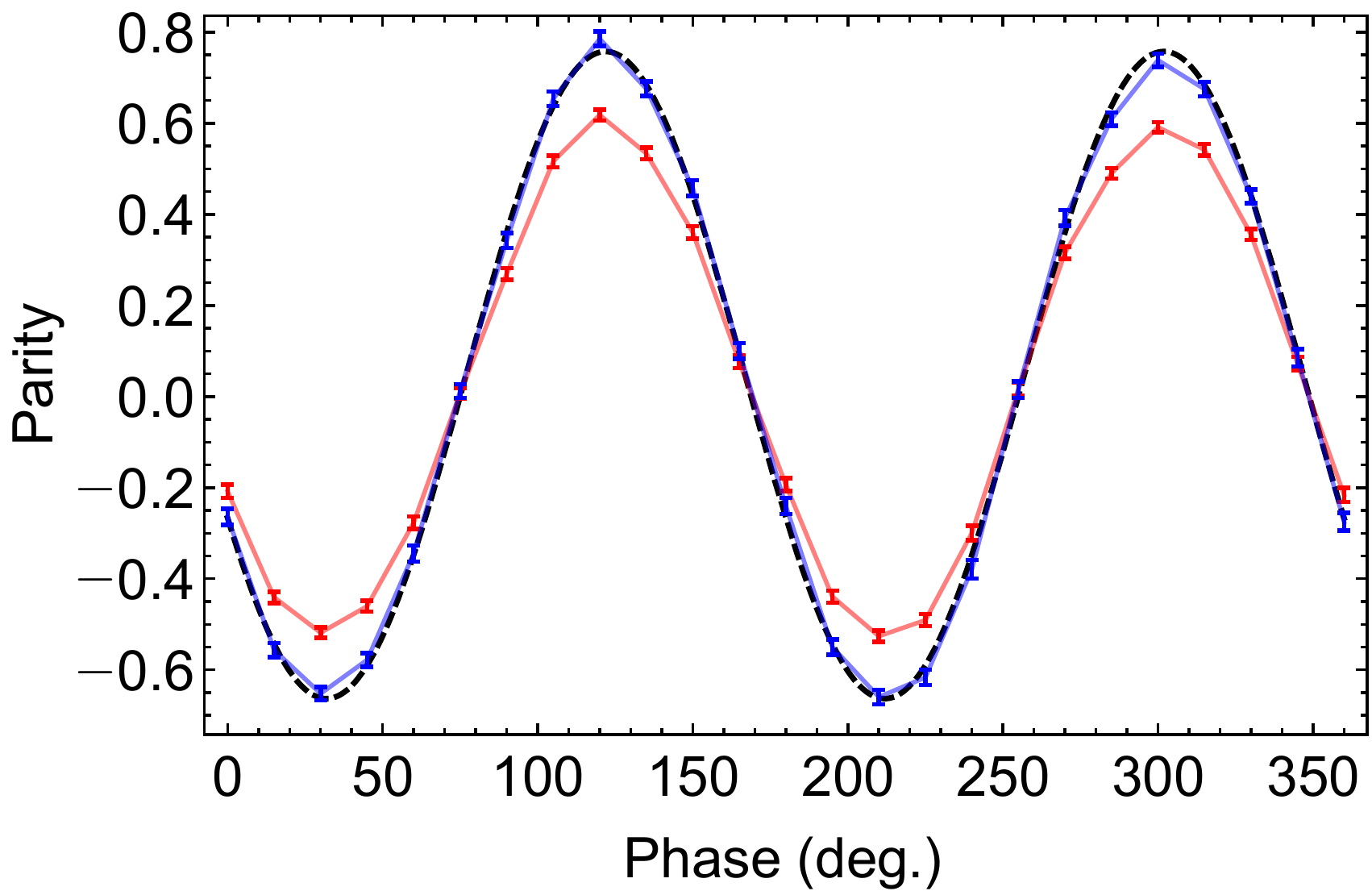}%
\caption{\label{Parity} Parity measurement for the maximally-entangled state. The blue data points show the measurement-error corrected data, whereas the red points are uncorrected for measurement errors. The dashed line is a sinusoidal fit two extract the magnitude of the entangled state coherences $|\rho_{00,11}|$ (see main text). } 
\end{figure}

We analyze the entangled state that is produced by applying a final $\pi/2$-pulse, about an axis that is rotated by an angle $\phi$ with respect to the rotation axes of the preceding $\pi/2$ and $\pi$ pulses. The expectation value of the parity operator defined as $\langle \hat{\Pi}\rangle =  P_{DD}+P_{BB} - P_{BD} -P_{DB}$, where the subscript indicates atom~1 and atom~2, respectively in either a bright ($B$) or dark ($D$) state, 
witnesses the degree of entanglement. The resulting parity is shown in Fig.~\ref{Parity}, from which we infer the coherences of density matrix of the prepared entangled state are $\vert 2 \rho_{11,00}\vert = 0.71(1)$ \cite{bollinger1996optimal,sackett2000experimental}. We measure the diagonal elements of the density matrix $\rho_{11,11}$ and $\rho_{00,00}$ by omitting the final  $\pi/2$-pulse and measuring $P_{BB}$ and $P_{DD}$. Because atoms in state $\vert1\rangle=\vert F=4, m_{F} =0\rangle$ are indistinguishable from atoms in $F=4$ with $m_{F}\neq 0$, the analysis of Ref.~\cite{sackett2000experimental} will lead to an over-estimate of fidelity when population can exist outside the qubit manifold because of imperfect optical pumping. We find that 
\begin{equation}
\mathcal{F}=\frac{1}{2}\left(P_{DD} + P_{BB}\right) + \frac{1}{2}\vert \rho_{11,00}\vert - \varepsilon_{\mathrm{OP}},
\end{equation}
where $2 \varepsilon_{\mathrm{OP}}= 1-P_{1, m_{F}=0} P_{2, m_{F}=0}$, and $P_{1, m_{F}=0}$ ($P_{2, m_{F}=0}$) is the probability that atom~1 (atom~2) is optically pumped into the desired $m_{F} = 0$ level. We find that 
$\mathcal{F}=0.77 \pm 1\times 10^{-2} \; \mathrm{(stat.)} \; \pm 3\times 10^{-2} \; \mathrm{(sys.)}$
, where the only correction applied thus far is to correct for state detection (\textit{i.e.}, measurement errors). When we account for both state-preparation and measurement (SPAM) errors, we find that 
$\mathcal{F}_{\mathrm{SPAM}} = \mathcal{F}/\left(P_{1, m_{F}=0} P_{2, m_{F}=0}\right) = 0.85\pm 1\times 10^{-2} \; \mathrm{(stat.)} \; \pm 3\times 10^{-2} \; \mathrm{(sys.)}$. 
Here, the systematic uncertainties result from propagating the systematic uncertainty in bright state detection efficiency. Hence, in the limit of perfect qubit state preparation and measurement, we infer that we would produce the desired output state with approximately 85\% fidelity.

Infidelity arising from atomic motion, inhomogeneous coupling to Rydberg state, and off-resonant scatter from intermediate levels is negligible and consistent with the analysis of Ref.~\cite{Mitra2020}.  For 10~$\mu$K temperatures, errors resulting from Doppler shifts are less than 0.1\% because the echo topology is highly effective at removing errors that are constant in time, such as detuning errors caused by the Doppler effect. In the absence of the echo, we find that for these detuning parameters, the gate error will be at the level of 2\%, which results directly from the uncompensated single-atom light shift Hamiltonian terms. Further, we find that errors arising from atom-position fluctuations that couple to $\phi_{J}$ are small and will also contribute at the 0.1\% level. We note that operating at larger inter-atomic separation, outside the blockade radius, can result in greater sensitivity to position fluctuations because of the large gradient of $J$ with respect to the vector connecting the two atoms (both its magnitude and angle relative to the quantization axis).

Other common imperfections are likewise suppressed in our approach.  The Rabi frequency homogeneity coupling the two atoms to the Rydberg level is ensured by aligning the coupling laser along the interatomic axis and using sufficiently large laser beam waist. This is enabled in our scheme by removing the requirement to individually address each atom.  Photon scattering from intermediate levels is highly-suppressed by using direct coupling to the Rydberg levels with an ultra-violet laser.  Remaining photon scattering arises from the Rydberg state lifetime.  For the parameters used in this work, the probability of spontaneous decay of the Rydberg state is approximately 0.25\% for $64P$ lifetime of 170~$\mu$s at a background blackbody radiation temperature of 300~K \cite{vsibalic2017arc}. Because the ramp and gate times scale inversely with $\Omega$, higher Rydberg laser power will reduce the probability of spontaneous emission. Further, using so-called ``shortcuts to adiabaticity'' \cite{Chen2010} may allow a faster ramp on resonance, which in turn allows a larger fraction of $\phi_{J}$ to accumulate during the static phase of the sequence. Using a combination of these techniques, we expect it is possible to reduce the probability of spontaneous decay in this approach by an order of magnitude. We further note that systems with larger oscillator strengths, such as the $^{3}P_{0}\rightarrow {}^{3}S$ Rydberg series found in alkaline earth atoms, will naturally enhance the gate speed.  

We attribute much the observed infidelity in our experiment to laser frequency noise.  The eigenvalues $E_{10}^{\pm}$, $E_{01}^{\pm}$, and $E_{11}^{\pm}$ all depend on laser detuning, and the echo topology relies on stationary detuning errors over the sequence for its effectiveness. Specifically, two main types of errors arise from laser noise: (1) errors in $\phi_{J}$ that arise from fluctuations in $J\left(\Delta, \Omega\right)$, and (2) imperfect cancellation of the single atom light shifts by the echo.  We find that the entangling operation has a sensitive dependence to the details of the laser noise spectrum. In terms of the rotation angles used in Ref.~\cite{Mitra2020}, Eqn.~\ref{UGate} can be extended to include rotation angle errors that arise from non-stationary laser frequency noise as
\begin{equation}
\hat{U}_{\mathrm{gate}} = e^{-i\left[\varphi_{1}\left(t_{1}\right)- \varphi_{1}\left(t_{2}\right)\right] \hat{S}_{y} -\frac{i}{2}\left[\varphi_{2}\left(t_{1}\right)+ \varphi_{2}\left(t_{2}\right)\right] \left(\hat{S}_{y}\right)^{2} }    
\end{equation}
where $\varphi_{1}\left(t_{0}\right) = \int_{t_{0}}^{t_{0}+t_{d}} \left[E_{\mathrm{LS}}^{(1)}\left(t\right)/\hbar+J\left(t\right)/2\right] dt$, and $\varphi_{2}\left(t_{0}\right) = \int_{t_{0}}^{t_{0}+t_{d}} J\left(t\right) dt$. Here, $t_{1}$ ($t_{2}$) corresponds to the first (second) dressing pulse turn-on time with duration $t_{d}$. Given lower bound estimates of the laser noise spectrum we calculate that laser noise will reduce the fidelity in the range of 5\% to 10\%.  Well-known challenges to accurate laser noise estimation as well as day-to-day fluctuations in its magnitude limit our ability to precisely quantify this effect with the current apparatus.  We plan a detailed analysis of this subject in follow-on work. 

We additionally estimate that sources gate error arising from combinations imperfect detection fidelity and loss of atoms during the entangling sequence are at the few percent level. Improvements in detection fidelity (\textit{e.g.}, Ref.~\cite{kwon2017parallel}), will greatly reduce these contributions. 

We experimentally demonstrate a M{\o}lmer-S{\o}rensen gate between two neutral atoms using single-photon dressing to Rydberg levels.  We show that this gate is highly immune to noise sources and experimental imperfections that have limited the performance of other approaches to neutral atom gates.  With further experimental refinement, we expect that high-fidelity gates as forecast in \cite{Mitra2020} can be achieved.  Our approach is a competitive technique for realizing complex many-body architectures for quantum information experiments exploring computation, simulation, and quantum metrology.

\begin{acknowledgments}
This work was supported by the Laboratory Directed Research and Development program at Sandia National Laboratories. Sandia National Laboratories is a multimission laboratory managed and operated by National Technology and Engineering Solutions of Sandia, LLC., a wholly owned subsidiary of Honeywell International, Inc., for the U.S. Department of Energy’s National Nuclear Security Administration under contract DE-NA-0003525. This paper describes objective technical results and analysis. Any subjective views or opinions that might be expressed in the paper do not necessarily represent the views of the U.S. Department of Energy or the United States Government This work was also supported by the Laboratory Directed Research and Development program of Los Alamos National Laboratory under project number 20190494ER.
\end{acknowledgments}



%

\end{document}